\documentclass[aps,prb,twocolumn,showpacs,superscriptaddress]{revtex4}
\usepackage{graphicx}
\usepackage{bm}
\newcommand{\beq}{\begin{equation}}
\newcommand{\eeq}{\end{equation}}

\begin{document}
\title{Spontaneous breaking of four-fold rotational symmetry
in two-dimensional electronic systems explained as a continuous
topological transition}

\author{M.~V.~Zverev}
\affiliation{Russian Research Centre Kurchatov
Institute, Moscow, 123182, Russia}
\author{J.~W.~Clark}
\affiliation{McDonnell Center for the Space Sciences \&
Department of Physics, Washington University,
St.~Louis, MO 63130, USA}
\author{Z.~Nussinov}
\affiliation{McDonnell Center for the Space Sciences \&
Department of Physics, Washington University,
St.~Louis, MO 63130, USA}
\author{V.~A.~Khodel}
\affiliation{Russian Research Centre Kurchatov
Institute, Moscow, 123182, Russia}
\affiliation{McDonnell Center for the Space Sciences \&
Department of Physics, Washington University,
St.~Louis, MO 63130, USA}
\date{\today}
\begin{abstract}
The Fermi liquid approach is applied to the problem of
spontaneous violation of the four-fold rotational point-group
symmetry ($C_4$) in strongly correlated two-dimensional
electronic systems on a square lattice.  The symmetry breaking
is traced to the existence of a topological phase transition.
This continuous transition is triggered when the Fermi line,
driven by the quasiparticle interactions, reaches the van
Hove saddle points, where the group velocity vanishes and the
density of states becomes singular.  An unconventional Fermi
liquid emerges beyond the implicated quantum critical point.

\end{abstract}

\maketitle

The breaking of fundamental symmetries in ground states of strongly correlated two-dimensional (2D) electron systems
\cite{campuzano,ando,kapitulnik,hinkov,mook,taillefer}
remains one of the most intensely debated topics in low-temperature condensed matter physics. Kivelson, Fradkin, and Emery \cite{nematic} were the first to discuss the case of nematic phase transitions, well before relevant experimental data was obtained. Somewhat later, Yamase and Kohno \cite{yamase1} (within $t-J$ model) and Halboth and Metzner \cite{metzner1} (within the Hubbard model) attributed the breaking of four-fold symmetry to violation of a Pomeranchuk stability condition \cite{pom} associated with antiferromagnetic fluctuations.

Subsequently, much theoretical work has been aimed at elucidating
salient features of this phenomenon, primarily within mean-field theory.\cite{oganesyan,kampf,metzner2,kee,metzner3,lamas,Fradkin_review}
It is instructive to recognize that the approach taken in these
efforts bears a striking resemblance to that employed by Belyaev fifty years ago to describe quadrupole deformation in atomic nuclei.\cite{belyaev} To determine the critical point at which the spherical shape becomes unstable and calculate the nuclear deformation beyond this point, he introduced an effective Hamiltonian with separable quadrupole-quadrupole ${\bf Q}_1{\bf Q}_2$ interaction. Analogously, for two-dimensional tetragonal electronic systems,
a separable interaction $d_2({\bf p}_1) d_2({\bf p}_2)$ with order parameter $d_2(p_x,p_y){=}\cos p_x{-}\cos p_y $ is adopted in the mean-field treatments of the breakdown of $C_4$ symmetry, momenta being measured in units of the inverse lattice constant. However, such an effective Hamiltonian with separable interaction is appropriate only in the channel where symmetry breaking occurs. Moreover, even in this channel a mean-field approach may be inadequate, as exhibited for example in the prediction of a first-order phase transition in the case of violation of point-group symmetries on a square lattice.\cite{khavkine}

Burdened with variety of inconsistencies, the mean-field description
of nuclear deformation was superseded many years ago by the more
sophisticated Fermi-liquid (FL) approach.\cite{migdal} Following this successful precedent, we work within the FL framework to obtain a better understanding of $C_4$-symmetry breaking in electron systems on a 2D square lattice. Intensive numerical calculations assuming a finite-range exchange interaction, supported by complementary analysis
of a simplified model, disclose unexpected features of the phenomenon.
In contrast to the description given by mean-field theory, we find
that the breakdown of $C_4$ symmetry is associated with a
{\it topological} phase transition that occurs under conditions
that allow the Fermi line, {\it calculated within FL theory},
to reach the van Hove saddle points $(0,\pi),(\pi,0),(0,-\pi),(-\pi,0)$.

Consideration  of topological transitions dates back to an article by
I.\ M.\ Lifshitz \cite{lifshitz}, in which the form of the
single-particle spectrum $\epsilon({\bf p})$ was assumed to be known.  However, within FL theory $\epsilon({\bf p})$ is itself a functional of the quasiparticle momentum distribution $n({\bf p})=\left[1+\exp\left({(\epsilon({\bf p}) -\mu)/T}\right)\right]^{-1}$. Accordingly, self-consistent inclusion of the interactions between
quasiparticles may lead to unforeseen types of the topological
transitions.\cite{ks,vol1,noz,zb,shagp,jetplett2001,schofield,haochen,volrev,shagrev,prb2008,jetplett2009}
We find just such a case in the problem of $C_4$-symmetry violation.

Stated simply, topological transitions in correlated Fermi systems are signaled (at zero temperature) by a change of the number of roots of equation
\beq
\epsilon({\bf p},n_F)=\mu,
\label{topeq}
\eeq
where $n_F$ is the Fermi step distribution and $\mu$ is the chemical potential. For a thorough development of the concept, see the review by Volovik.\cite{volrev} Throughout, we adhere to his rigorous quantitative definition of topological phase transitions, as distinguished from looser notions such as transitions between large and small Fermi surfaces that are also prevalent in the literature.

Analysis of topological phase transitions in fermionic systems
is greatly facilitated by the absence of {\it critical} fluctuations of any order parameter at the transition point and its vicinity, meaning that the Landau-Migdal quasiparticle picture retains its validity. Thus, the physical many-fermion system may be viewed as a system of interacting quasiparticles, and $C_4$ symmetry violation can be investigated using the fundamental FL relation \cite{lan,lanl}
\beq
{\partial\epsilon({\bf p})\over \partial {\bf p}}=
{\partial\epsilon_0({\bf p})\over \partial {\bf  p}} +{1\over 2}{\rm Tr}
\int {\cal F}_{\alpha\beta,\alpha\beta}({\bf p},{\bf p}_1)
{\partial n({\bf p}_1)\over \partial {\bf p}_1} d\upsilon_1.
\label{lansp}
\eeq
In this relation, $d\upsilon= dp_xdp_y/(2\pi)^2$ is the volume element of 2D momentum space, $\epsilon_0({\bf p})$ is the bare single-particle spectrum, and ${\cal F}$ is a phenomenological interaction function depending only on the {\it momenta} ${\bf p}$ and ${\bf p}_1$ of the colliding quasiparticles.

Our goal is to analyze the impact of antiferromagnetic fluctuations on electron spectra calculated using Eq.~(\ref{lansp}).  Taking account of these fluctuations in the interaction function ${\cal F}$ presents little difficulty in the regime {\it far} from the  antiferromagnetic phase transition, since the fluctuation exchange can be treated within the Ornstein-Zernike approximation. The part of ${\cal F}$ responsible for the exchange is then given by
\beq
{\cal F}^e_{\alpha\beta\gamma\delta}({\bf p},{\bf p}_1)=
\lambda^2 {\bm \sigma}_{\alpha\beta}{\bm \sigma}_{\gamma\delta}
\left[({\bf p}{-}{\bf p}_1{-}{\bf Q})^2+\xi^{-2}\right]^{-1},
\label{sfl}
\eeq
where the constant $\lambda$ represents the spin-fluctuation vertex and ${\bf Q}=(\pi,\pi)$ the antiferromagnetic wave vector, while $\xi$ denotes the correlation radius. Result (\ref{sfl}) relies on the fact that the interaction function ${\cal F}$ coincides with a specific static limit of the quasiparticle scattering amplitude whose initial and final energies are on the Fermi surface, such that this quantity is {\it energy- and frequency-independent}.\cite{lan,lanl}

Inserting Eq.~(\ref{sfl}) into Eq.~(\ref{lansp}) and evaluating the spin-fluctuation contribution aided by the identity $2{\bm \sigma}_{\alpha\beta}{\bm \sigma}_{\gamma\delta}=
3\delta_{\alpha\delta}\delta_{\gamma\beta}- {\bm \sigma}_{\alpha\delta} {\bm \sigma}_{\gamma\beta}$, one arrives at
\beq
\epsilon({\bf p})=\epsilon_0({\bf p}) +f_a
\int { n({\bf p}_1)\over ({\bf p}-{\bf p}_1-{\bf Q})^2+\xi^{-2}}\,
d\upsilon_1,
\label{spec}
\eeq
where $f_a=3\lambda^2/2$. The chemical potential $\mu$, being constant along the Fermi line, is determined by the normalization condition $\int n({\bf p})\,2d\upsilon=\rho$, where factor 2 assumes summation over two spin projections.

Numerical solution of the 2D nonlinear integral equation (\ref{spec})
is extremely time-consuming since it is necessary to compute to high precision to rule out spurious signals of broken symmetry.

Calculations have been performed in the case of an open Fermi surface, assuming a tight-binding spectrum
\beq
\epsilon_0({\bf p}) = -2\,t_0\,(\cos p_x + \cos p_y)
+4\,t_1\,\cos p_x\cos p_y-\mu,
\label{tight}
\eeq
with the ratio $t_1/t_0$ of input parameters $t_0$ and $t_1$ taken
to ensure a rather small distance between the tight-binding Fermi line and the saddle points. A finite-range interaction function
\beq
f({\bf q})=f_a\left[({\bf q}-{\bf Q})^2+\xi^{-2}\right]^{-1}
\label{finite}
\eeq
is adopted, with $\xi=30$. Salient results are reported for
strategically chosen values $f_a = 1.0$ and $1.5$ of the coupling parameter (in units of $2t_0$), and at temperatures
$T = 10^{-2}$ and $10^{-4}$ (also in units of $2t_0$).

\begin{figure}[t]
\includegraphics[width=0.73\linewidth,height=1.\linewidth]{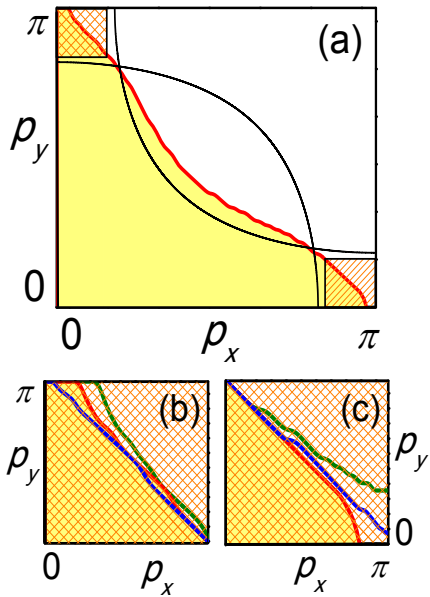}
\caption{
Fermi lines computed for the Fermi-liquid-theory model assuming bare
tight-binding spectrum (\ref{tight}) with $t_1/t_0=0.45$ and
finite range interaction (\ref{finite}) with $\xi = 30$.
Panel (a): Results for $f_a=1.5$ and $T=10^{-4}$ (both in units of
$2\,t_0$). Thick solid line (in red): one of two identical solutions with spontaneously broken $C_4$ symmetry. Only the first quadrant of the Brillouin zone is drawn, since neither $p_x\to -p_x$ nor $p_y\to -p_y$ reflection symmetry is broken. Thin solid lines (in black): Fermi lines for the bare tight-binding spectrum $\epsilon^0_{\bf p}$ and its counterpart. Panels (b) and (c): Two shaded squares adjacent to the saddle points $(0,\pi)$ and $(\pi,0)$ present in panel (a) are magnified. The Fermi-line solution with broken $C_4$ symmetry appearing in panel (a) (red line) is drawn together with two $x-y$-symmetrical solutions corresponding respectively to $f_a=1.0$ and $T=10^{-4}$ (green line), and to $f_a=1.5$ and $T=10^{-2}$ (blue line).
}
\label{fig:xyfinite_range}
\end{figure}

\begin{figure}[t]
\includegraphics[width=0.9\linewidth,height=0.75\linewidth]{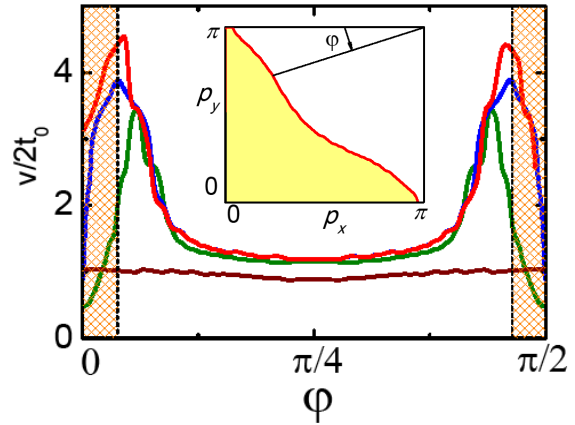}
\caption{
Group-velocity magnitudes $v=|\partial\epsilon({\bf p})/\partial {\bf p}|$ (in units of $2t_0$), evaluated along the Fermi line as a function of the angle $\varphi$ defined in the inset, for different single-particle spectra $\epsilon({\bf p})$. Results
are shown for the bare tight-binding model with the same parameter
choice as in Fig.~\ref{fig:xyfinite_range} (brown line) and for the Fermi-liquid-theory model of Fig.~\ref{fig:xyfinite_range} at $f_a=1.0$, $T=10^{-4}$ (green line);  $f_a=1.5$, $T=10^{-4}$ (red line); and $f_a=1.5$, $T=10^{-2}$ (blue line). Broken $C_4$ symmetry of the red curve with respect to $x-y$ exchange is manifested by its different behavior in the two shaded areas close to the saddle points.
}
\label{fig:xyvelo}
\end{figure}

\begin{figure}[t]
\includegraphics[width=0.9\linewidth,height=1.0\linewidth]{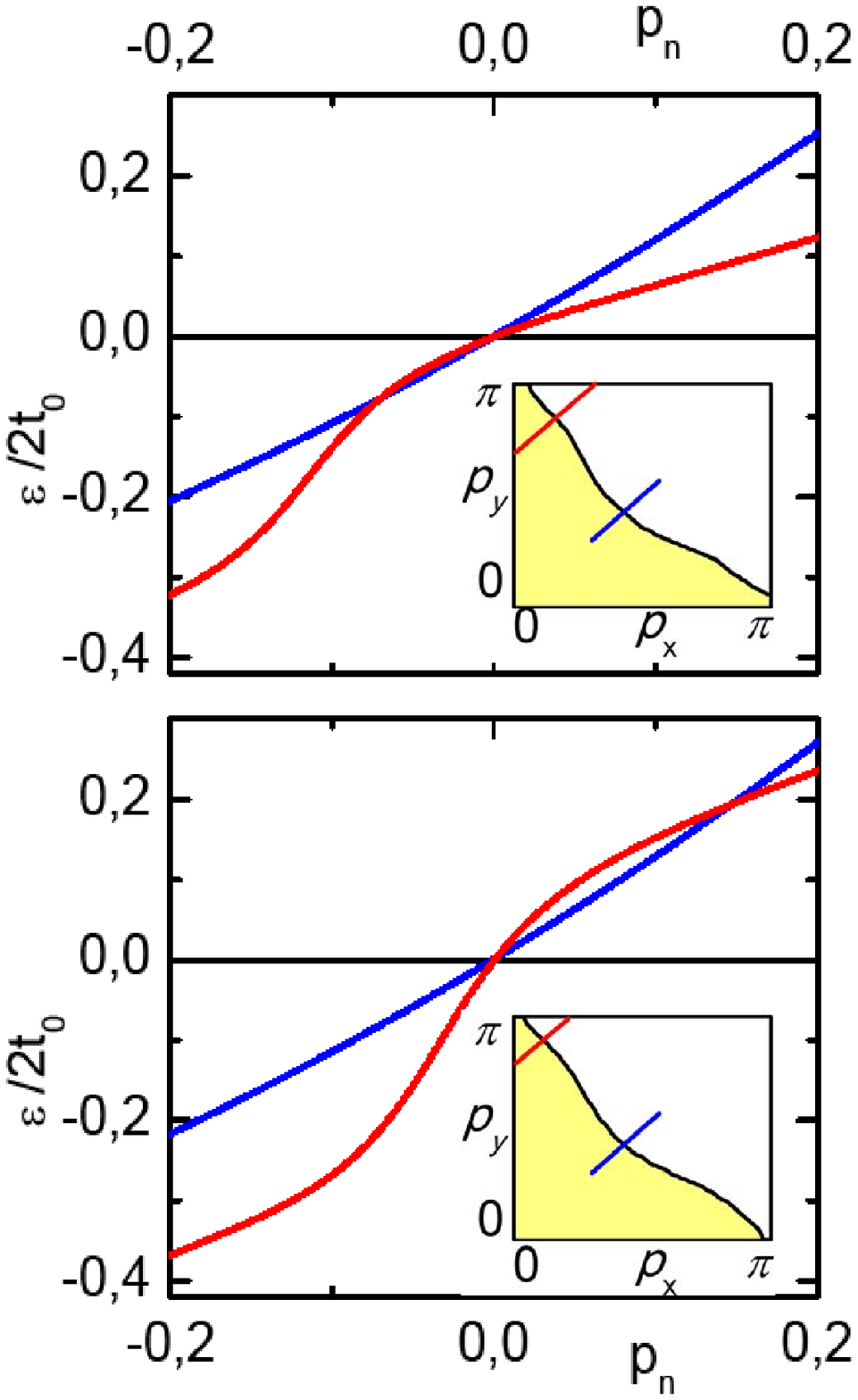}
\caption{
Single-particle spectra $\epsilon(p)$ (in units of $2t_0$) evaluated
along two lines in the momentum plane (as indicated with corresponding color coding in the insets). The Fermi-liquid theory model used for Fig.~\ref{fig:xyfinite_range} is applied at $T{=}10^{-4}$ with $f_a{=}1.0$ (upper panel) and $f_a{=}1.5$ (lower panel). Spectral curves are plotted versus the normal component $p_n$ of the momentum, measured with respect to the Fermi line. In the insets, the blue line coincides with the diagonal of the zone, while the red line crosses the relevant hot line.
}
\label{fig:xyspectra}
\end{figure}

The numerical calculations, as represented here in
Figs.~\ref{fig:xyfinite_range}--\ref{fig:xyspectra},
reveal some remarkable features of the quasiparticle rearrangement
responsible for violation of $C_4$ symmetry. One conspicuous
feature, well documented in the figures, is that only those
quasiparticles residing in domains close to the saddle points are noticeably affected by the inclusion of antiferromagnetic fluctuations in the FL treatment.

To understand of the onset of symmetry breaking, it is most instructive to track the distance between neighboring points where the Fermi line crosses the border of the Brillouin zone. This distance is found to shrink as the coupling constant $f_a$ is increased toward a critical value lying between $1.0$ and $1.5$. Breakdown of $C_4$ symmetry presumably occurs for a critical coupling $f_{ac}$ at which the two crossing points {\it merge} with one another as they embrace and converge upon the nearby saddle point. As seen in Fig.~\ref{fig:xyfinite_range}, the Fermi line (in red)
calculated at temperature $T=10^{-4}$ (effectively zero) for a coupling constant $f_a=1.5$ lying beyond the critical point does indeed violate $x-y$ symmetry; the corresponding Fermi line for $f_a=1.0$ (in green) does not. Another feature worthy of note is the effect of temperature in suppressing the symmetry breaking phenomenon.
Comparison of the deviations of the Fermi line calculated at $f_a=1.5$
and $T=10^{-4}$ (in red) from the symmetry-preserving Fermi line obtained at $f_a=1.5$ and $T=10^{-2}$ (in blue) demonstrates that $C_4$ symmetry can be restored by elevating the temperature.

%%%%%%%%%%%%%%%%%%%%%%%%%%%%%%%%%%%%%%%%%%%%%%%%%%%%%%%%%%%%%%%
%%%%%                Figure: xyfinite_range.eps
%%%%%%%%%%%%%%%%%%%%%%%%%%%%%%%%%%%%%%%%%%%%%%%%%%%%%%%%%%%%%%%
%%%%%%%%%%%%%%%%%%%%%%%%%%%%%%%%%%%%%%%%%%%%%%%%%%%%%%%%%%%%%%%

%%%%%%%%%%%%%%%%%%%%%%%%%%%%%%%%%%%%%%%%%%%%%%%%%%%%%%%%%%%%%%%
%%%%%                Figure: xyvelo.eps
%%%%%%%%%%%%%%%%%%%%%%%%%%%%%%%%%%%%%%%%%%%%%%%%%%%%%%%%%%%%%%%
%%%%%%%%%%%%%%%%%%%%%%%%%%%%%%%%%%%%%%%%%%%%%%%%%%%%%%%%%%%%%%%%%

Results from calculations of the magnitude
$v({\bf p})=|\partial\epsilon({\bf p})/\partial {\bf p}|$ of the group velocity along the Fermi line are plotted in Fig.~\ref{fig:xyvelo}. These results demonstrate that the impact of antiferromagnetic correlations, as described by Eq.~(\ref{sfl}), is only significant for quasiparticles in momentum domains adjacent to the saddle points. The small gap between the values for $v$ given by the bare tight-binding model and by the Fermi-liquid-theory treatment, seen in domains away from the saddle points, is due primarily to a shifting of the location of the Fermi line triggered by the antiferromagnetic correlations. On the other hand, the group velocity $v$ evaluated at $f_a=1.5$ and $T=10^{-4}$ (red line) is seen to acquire an $x-y$-anisotropic component close to the saddle points.

%%%%%%%%%%%%%%%%%%%%%%%%%%%%%%%%%%%%%%%%%%%%%%%%%%%%%%%%%%%%%%%
%%%%%                Figure: xyspectra.eps
%%%%%%%%%%%%%%%%%%%%%%%%%%%%%%%%%%%%%%%%%%%%%%%%%%%%%%%%%%%%%%%
%%%%%%%%%%%%%%%%%%%%%%%%%%%%%%%%%%%%%%%%%%%%%%%%%%%%%%%%%%%%%%%

Fig.~\ref{fig:xyspectra} presents results from calculations of electron spectra at $T=10^{-4}$ in the direction perpendicular to the Fermi line. Far from the saddle points, the impact of antiferromagnetic fluctuations is again found to be minor, but in their vicinity the effects are very strong. In particular, the particle and hole spectra cease to be alike; the average slope of the hole spectrum noticeably exceeds that of the particle spectrum. One might attribute this difference to the variation of Fermi-line contributions to Eq.~(\ref{lansp}) associated with a turning point emerging in the trajectory of the Fermi line at the critical point. From all the results discussed above, we infer that the description of the rearrangement of the ground state in terms of a single $d_2$ parameter is a poor approximation.

To further analyze and interpret the results obtained numerically for
the finite-range interaction (\ref{finite}) and bare tight-binding
spectrum (\ref{tight}), we simplify the interaction function in the
manner of Ref.~\onlinecite{jetplett2001}, replacing the exchange
term (\ref{sfl}) by an infinite-range form
\beq
f({\bf q}) = (2\pi)^2 f_0\,\delta({\bf q}-{\bf Q}),
\label{infinite}
\eeq
with coupling constant $f_0$. Eq.~(\ref{spec}) is then replaced
by \cite{jetplett2001}
\beq
\epsilon({\bf p},T)=\epsilon_0({\bf p})+f_0n(\epsilon({\bf p}+{\bf Q},T))
\label{noza}.
\eeq
This treatment is analogous to that adopted by Nozi\`eres \cite{noz} in a study of non-FL behavior of strongly correlated Fermi systems for the case where forward scattering is dominant. Eq.~(\ref{noza}) can be derived within a standard variational procedure based on the formula \cite{jetplett2001}
\beq
E=\int \left[\epsilon^0_{\bf p}n({\bf p})
  +{1\over 2}f_0n({\bf p})n({\bf p}+{\bf Q})\right]\, 2d\upsilon
\label{enan}
\eeq
for the energy $E$ of the model quasiparticle system. Eq.~(\ref{noza}) is conveniently rewritten as a system of two equations
\begin{eqnarray}
  \epsilon_1&=&\epsilon^0_1+f_0 n(\epsilon_2),\nonumber\\
  \epsilon_2&=&\epsilon^0_2+f_0 n(\epsilon_1),
  \label{set1}
\end{eqnarray}
where  $\epsilon_1=\epsilon({\bf p}_1)-\mu$ and
$\epsilon_2=\epsilon({\bf p}_1+{\bf Q})-\mu$, while
$\epsilon^0_1=\epsilon_0({\bf p})$ and
$\epsilon^0_2=\epsilon_0({\bf p}+{\bf Q})$.

In the earlier work,\cite{jetplett2001} a graphical procedure was
introduced to solve the set (\ref{set1}) at $T=0$. Three different solutions were found. One of these corresponds to an exceptional, non-FL state \cite{prb2008} exhibiting a flat single-particle spectrum. In the absence of pairing correlations, this solution turns out to be disfavored energetically relative to the other two solutions, which possess identical FL-like properties.

Focusing on the properties of the latter two solutions, we observe that at $T=0$ the associated rearrangement of the initial Landau state can occur only in those 2D systems in which hot spots \cite{pines} exist---points situated on the Fermi line and connected by the vector ${\bf Q}$. In fact, for systems with small quasiparticle filling, the product $n({\bf p})n({\bf p}+{\bf Q})$ vanishes for any momentum
${\bf p}$; hence the ground-state energy is independent of the coupling constant $f_0$. The same is true in the case of small quasihole filling.

In systems having hot spots, the rearrangement occurs due to breaking of quasiparticle pairs occupying single-particle states with momenta ${\bf p}$ and ${\bf p}+{\bf Q}$. The corresponding domain ${\cal R}$ (the ``reservoir'') consists of four quasi-rectangles, each adjacent to one of the van Hove saddle points. Each of the four elements of ${\cal R}$ is confined between the border of the Brillouin zone, the {\it counterpart} of the initial Fermi line, defined by the equation $\epsilon_0({\bf p}+{\bf Q})=\mu$, and two segments of the Fermi line embracing the given saddle point.

\begin{figure}[t]
\includegraphics[width=0.8\linewidth,height=1.1\linewidth]{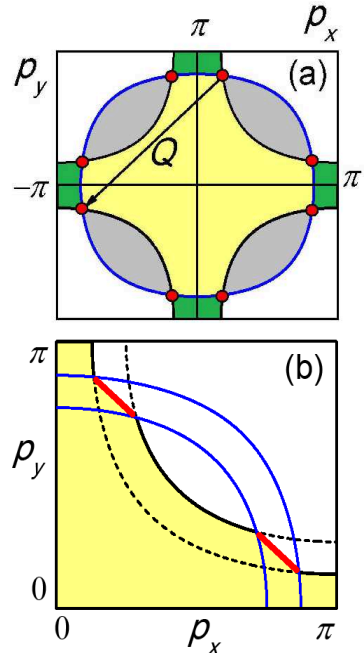}
\caption{
Panel (a): Fermi line (black) and its counterpart (blue) for the bare tight-binding spectrum of Eq.~(\ref{tight}) with $t_1/t_0=0.45$. The ``reservoirs'' ${\cal R}$ (see text) are colored in green, and the lenses ${\cal L}$, in light gray. The hot spots connected with each other by the vector ${\bf Q}$ are symbolized by red dots. Panel (b): Fermi line for the simplified Fermi-liquid-theory model based on the infinite-range interaction function (\ref{infinite}) with $f_0{=}0.4$ (in units of $2t_0$). Hot lines are drawn as red line. Fermi lines for the bare tight-binding spectrum $\epsilon_0({\bf p})$ and for the same spectrum shifted by $-f_0$ are shown as dashed lines while their counterparts are drawn in blue.
}
\label{fig:xyzero}
\end{figure}

In the rearrangement being considered, the quasiparticles move out
the domain ${\cal R}$ to resettle in a region ${\cal L}$ where all pairs of single-particle states connected by the vector ${\bf Q}$ are empty. The region ${\cal L}$ comprises four ``lenses,'' situated between neighboring hot spots and bounded by the initial Fermi line
and its counterpart (see panel (a) of Fig.~\ref{fig:xyzero}). The transfer of one quasiparticle from ${\cal R}$ to ${\cal L}$ produces a gain in energy which is just the coupling constant $f_0$ minus the loss $\tau$ of kinetic energy. The minimum loss $\tau_{\rm min}$ occurs when a quasiparticle, vacating a state in ${\cal R}$ with momentum ${\bf p}$, occupies in ${\cal L}$ a state of lowest energy, given by the chemical potential, so that $\tau_{\rm min}= \mu - \epsilon_0({\bf p})$. Therefore the rearrangement is favorable provided
$\epsilon_0({\bf p})-\mu+ f_0\geq 0$. In the resettlement process,
the chemical potential $\mu$, which coincides with the maximum
quasiparticle energy in occupied states (in particular, in the lens
region), evidently increases relative to its initial value $\mu_i$.
The quasiparticles that resettle to the lens region then possess almost the same effective mass as the noninteracting electrons. This conclusion is confirmed by the numerical calculations represented in Fig.~\ref{fig:xyspectra}.

%%%%%%%%%%%%%%%%%%%%%%%%%%%%%%%%%%%%%%%%%%%%%%%%%%%%%%%%%%%%%%%
%%%%%               Figure: xyzero.eps
%%%%%%%%%%%%%%%%%%%%%%%%%%%%%%%%%%%%%%%%%%%%%%%%%%%%%%%%%%%%%%%
%%%%%%%%%%%%%%%%%%%%%%%%%%%%%%%%%%%%%%%%%%%%%%%%%%%%%%%%%%%%%%%%%%%%%%%%

An alternative process involves transfer of the quasiparticle
counterpart, which has momentum ${\bf p}+{\bf Q}$. In this case, the
rearrangement occurs provided $ \epsilon_0({\bf p}+{\bf Q})-\mu+f\geq 0$. The choice between the two options is decided by comparison of the
corresponding energies. The boundary at which one behavior gives way to the other is defined by the relation $\epsilon_0({\bf p})=\epsilon_0({\bf p}+{\bf Q})$. Since the straight line so defined is part of the {\it new} Fermi line, we infer that the rearrangement has converted the original, isolated hot spot into a {\it continuous line
of hot spots} (see panel (b) of Fig.~\ref{fig:xyzero}).

The results obtained imply that quasiparticles are swept from a certain subdomain ${\cal S}$ of ${\cal R}$ consisting of eight approximately trapezoidal strips. The boundaries of a given strip are traced on three sides by (respectively) the initial Fermi line, the border of the Brillouin zone, and a line geometrically similar to the initial Fermi line but shifted into the domain ${\cal R}$ (see Fig.~\ref{fig:xyzero}). The strip's fourth side (red line) is just the {\it hot line}. The solution derived is self-consistent: any single-particle state with momentum ${\bf p}\in {\cal S}$ has its counterpart, with momentum ${\bf p}+{\bf Q}$, located outside ${\cal S}$, and this state is occupied, so that Eq.~(\ref{noza}) is fulfilled.  Transparently, in this non-critical scenario, the new momentum distribution {\it does not} violate $C_4$ symmetry.

In the situation where $C_4$ symmetry is violated in the rearrangement, the symmetry breaking occurs for a critical value $f_c$ of $f_0$, at which two segments of the Fermi line crossing the same boundary of the Brillouin zone {\it merge} at the saddle point.
When this happens, the number of solutions of Eq.~(\ref{topeq})
certainly drops, signaling a {\it topological phase transition} which, as readily seen, entails the breakdown of $C_4$ symmetry.

Suppose on the contrary that $C_4$ symmetry is preserved at $f_0>f_c$.  Then all the saddle points must be emptied simultaneously, implying that every rearranged saddle point energy $\epsilon_s$ {\it exceeds} the chemical potential $\mu$. But according to Eq.~(\ref{noza}), the interaction contribution to $\epsilon_s$ vanishes when all the saddle points are emptied. Consequently, at $f_0>f_c$, the saddle-point energy $\epsilon_s$ must coincide with the corresponding bare value $\epsilon^0_s$, which is {\it lower} than the initial chemical potential $\mu_i$. Thus, a contradiction is encountered.

We are driven to the conclusion that the critical situation giving rise to violation of $C_4$ symmetry is one in which the Fermi line, calculated within FL theory, attains a saddle point. Since both components of the quasiparticle group velocity ${\bf v}({\bf p}_s)$
vanish at this critical point, the corresponding density of states must acquire a singularity, which implies that we are dealing with a {\it quantum critical point} (QCP).

The contradiction is resolved beyond the QCP if only {\it one} of two neighboring saddle points is emptied, with the second remaining occupied---thereby breaking $C_4$ symmetry. As a point where the Fermi line crosses the $p_x$ axis moves away from the affected saddle point,
its counterpart, shifted by the vector ${\bf Q}$, slides along the border of the Brillouin zone, determining the boundary of the new filling. These conclusions drawn from analysis of the simple infinite-range model are in agreement the findings of the numerical
calculations based on the more elaborate model based on
Eqs.~(\ref{spec})--(\ref{finite}).

To summarize: in addressing the problem of $C_4$-symmetry violation, we have taken account of antiferromagnetic fluctuations within a self-consistent Fermi liquid approach, employing an interaction function that is more realistic than the separable approximation assumed in mean-field treatments.  We have demonstrated that inclusion of the exchange interaction drives the calculated single-particle spectrum so as to shrink the distance between saddle points and the Fermi line.  When merging occurs, the electron group velocity vanishes at the points of mergence, because these points coincide with the saddle points. A quantum critical point (QCP) of a new type is thereby revealed, at which a topological phase transition triggers the violation of $C_4$ symmetry. Significantly, the transition is found to
be {\it continuous}, in contrast to the first-order phase transition obtained in mean-field theory, where the corresponding QCP does not exist.
Beyond the transition point, the group velocity becomes finite again. Thus, on one side of the QCP, the system behaves as conventional Landau Fermi liquid.  On the other side, the electron liquid becomes an unconventional Fermi liquid because of the loss of four-fold symmetry.

We thank A.~Alexandrov, A.~Balatsky, E.~Fradkin, A.~Mackenzie,
V.~Shaginyan, V.~Yakovenko, and H.~Yamase for fruitful discussions.
This research was supported by the McDonnell Center for the Space Sciences, by Grants Nos.\ 2.1.1/4540 and NS-7235-2010.2 from the Russian Ministry of Education and Science, and by Grant No.~09-02-01284 from the Russian Foundation for Basic Research.

\end{document}